\documentclass{llncs}

\usepackage{epsfig}
 
\usepackage{listings}
\usepackage{graphicx}
\usepackage[all]{xy}
\usepackage{eepic,latexsym,amssymb}
\usepackage{fancybox,epic}
\usepackage{amsmath}

\usepackage{alltt}

\newcommand{\adom}{{\it ADom}}
\newcommand{\aint}{{\it AAssert}}
\newcommand{\aatom}{{\it AAtom}}
\newcommand{\acert}{{\it AApprox}}
\newcommand{\dom}{D_\alpha}
\newcommand{\inten}{Assert_\alpha}
\newcommand{\atom}{S_\alpha}
\newcommand{\cert}{Approx_\alpha}

\newcommand{\bicolano}{Bicolano}
\newcommand{\states}{State_{JVM}}
\newcommand{\interpreterp}{\mbox{\sc \jvmlb\_int}}

\newcommand{\lsem}{\mbox{$\lbrack\hspace{-0.3ex}\lbrack$}}
\newcommand{\rsem}{\mbox{$\rbrack\hspace{-0.3ex}\rbrack$}}
\newcommand{\sqbrack}[1]{\lsem #1 \rsem}

\newcommand{\short}[1]{}
\newcommand{\mycomment}[1]{\textbf{*** #1 ***}}

\long\def\comment#1{}

\newcommand{\jvmlb}{\textsc{jvml}$_{r}$}
\newcommand{\Jvmlb}{\textsc{jvml}$_{r}$ }
\newcommand{\Jvml}{\textsc{jvml} }
\newcommand{\jvml}{\textsc{jvml}}
\newcommand{\clp}{{LP}}

\newcommand{\MIS}{\textsc{mis}}

\newcommand{\class}{C}
\newcommand{\program}{P}
\newcommand{\interpreter}{\interpreterp}

\newcommand{\classreader}{\textsc{class\_reader}}
\newcommand{\specialization}{I_{P}}
\newcommand{\pe}{\textsc{partial\_e\-va\-lua\-tor}}
\newcommand{\verifier}{\textsc{ai\_verifier}}
\newcommand{\dotclass}{{\tt .class}}

\newcommand{\classes}{Class}
\newcommand{\programs}{\textit{Prog}}
\newcommand{\data}{\textit{Data}}
\newcommand{\jvmlp}{\jvmlb\_\programs}

\newcommand{\ciao}{\texttt{Ciao}}
\newcommand{\ciaopp}{\texttt{CiaoPP}}

\newcommand{\secbeg}{\vspace*{0pt}}
\newcommand{\secend}{\vspace*{0pt}}

\newcommand{\subsecbeg}{\vspace*{0pt}}
\newcommand{\subsecend}{\vspace*{0pt}}
\newcommand{\allttbeg}{\vspace*{-0.15cm}}
\newcommand{\allttend}{\vspace*{-0.15cm}}

\comment{\setlength{\textwidth}{12.8cm}
\setlength{\textheight}{21.1cm}

\setlength{\topmargin}{0cm}
\setlength{\oddsidemargin}{1.5cm}
\setlength{\evensidemargin}{1.5cm}
\setlength{\marginparwidth}{0.25cm}
\setlength{\marginparsep}{0.25cm}
}
\renewcommand{\intextsep}{2pt}

\title{Verification of Java Bytecode  using Analysis and
  Transformation of Logic Programs}

\author{E.~Albert\inst{1} \and M.~G\'omez-Zamalloa\inst{1}
  \and L.~Hubert\inst{2} \and G.~Puebla\inst{2}}

\institute{DSIC, Complutense University of Madrid, E-28040 Madrid, Spain\\
\and
CLIP, Technical University of Madrid, E-28660 Boadilla del Monte, Madrid, Spain \\
\email{\{elvira,mzamalloa,laurent,german\}@clip.dia.fi.upm.es}
}
\begin{document}
\maketitle

\vspace*{-0.4cm}

\begin{abstract}
  State of the art analyzers in the Logic Programming (LP) para\-digm
  are nowadays
  mature and sophisticated.
  They allow inferring a wide variety of global properties
  including termination, bounds on resource
  consumption, etc.  
  The aim of this work is to automatically transfer the power of such
  analysis tools for \clp\ to the analysis and verification of Java
  bytecode (\jvml).
  In order to achieve our goal, we rely on well-known techniques for
  meta-pro\-gram\-ming and program specialization.
  More precisely, we propose to partially evaluate a \jvml{}
  interpreter implemented in \clp\ together with (an \clp\
  representation of) a \jvml{} program and then analyze
  the residual program.
  Interestingly, at least for the examples we have studied, our
  approach produces very simple \clp\ representations of the original
  \jvml{} programs.  This can be seen as a
  decompilation from \jvml{} to high-level \clp\ 
  source. 
 By reasoning about 
such residual programs, we can automatically prove in the \ciaopp\ system some
    non-trivial properties of \jvml{} programs such as
    termination, run-time error freeness and infer bounds on
  its  resource consumption.
We are not aware of any other
  system which is able to verify such advanced properties of Java
  bytecode. 
\end{abstract}



\secbeg
\section{Introduction}
\secend
 
\short{The technique of \emph{abstract
interpretation} \cite{Cousot77-short} has
allowed the development of very sophisticated global static program
analyses which are at the same time automatic, provably correct, and
practical. The basic idea of \emph{abstract
interpretation} is to infer
information on programs by interpreting (``running'') them using
abstract values rather than concrete ones, thus, obtaining safe
approximations of programs behavior. 
\short{\emph{Abstract
interpretation} is
especially suitable for properties for which it is difficult to find
an expressive and syntactically decidable type system, but
nevertheless it is possible to obtain statically decidable, sufficient
conditions for the properties to hold. For such cases, abstract
domains are devised whose results are safe, though possibly incomplete
in the sense that, the fact that the analyzer fails to prove a
property should not be interpreted as that the property does not hold,
but rather that it may be due to the loss of precision, i.e.,
approximation, introduced by the abstraction. 
}
A classical application of the semantic approximations produced by an
abstract interpreter is to perform program \emph{verification}.
}

Verifying programs in the (Constraint) Logic Programming 
paradigm ---(C)LP---
offers a good number of advantages, an important one being the
maturity and sophistication of the analysis tools available for it.
The work presented in this paper is motivated by the existence of
\emph{abstract interpretation}-based analyzers~\cite{Cousot77-short}
which infer information on programs by interpreting (``running'') them
using abstract values rather than concrete ones, thus, obtaining safe
approximations of programs behavior.  These analyzers are parametric
w.r.t. the so-called abstract domain, which provides a finite
representation of possibly infinite sets of values. Different domains
capture different properties of the program with different levels of
precision and at different computational costs.  This includes error
freeness, 
data structure %
shape (like pointer sharing), bounds on data structure sizes, and
other operational variable instantiation properties, as well as
procedure-level properties such as determinacy, termination,
non-failure, and bounds on resource consumption (time or space cost),
etc.  \ciaopp~\cite{ciaopp-sas03-journal-scp-short-short} is the \emph{abstract
interpretation}-based preprocessor of the \ciao~(C)\clp{} system,
where analysis results have been applied to perform high- and
low-level optimizations and \emph{program verification}.

\begin{figure}[t]
\centering
\includegraphics[width=1.0\textwidth]{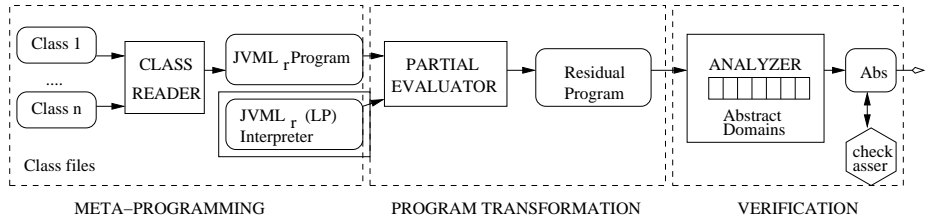}
\caption{Verification of Java Bytecode using Logic Programming
Tools}
\label{fig:acc-jvm}
\end{figure}

A principal advantage of verifying programs on the (\clp{})
\emph{source} code level is that we can infer  complex global
properties (like the aforementioned ones) for them. 
However, in certain applications like within the context of mobile
code, one may only have the \emph{object} code available.\short{, since mobile
components are typically deployed as bytecode. } In general, analysis
tools for such low-level languages are unavoidably more complicated
than for high-level languages because they have to cope with
complicated and unstructured control flow. Furthermore, as the \Jvml 
(Java Virtual Machine Language, i.e., Java bytecode) 
is a stack-based language, stacks cells are used to store
intermediate values, and therefore their type can change from one
assignment to another, and they can also be used to store 32 bits of a
64 bit value, which make the inference of stack information much more
difficult.
Besides, it is a non trivial task
to specify/infer global properties for the bytecode by using pre- and
post-conditions (as it is usually done in existing tools for
high-level languages).

The aim of this work is to provide a practical framework for the
verification of \Jvml which exploits the expressiveness,
automation and genericity of the advanced analysis tools for \clp\
source.\short{ The reasons why we choose Java Virtual (JVM) based
  environments are twofold.  First, they provide basic but important
  guarantees on the safety of incoming components.  In particular,
  virtual machines abstract away from many low-level details of the
  hardware processor.  As a result, the operations an abstract machine
  performs are more secure than their lower-level counterpart.
  Second, research prototypes of extensible JVMs are available for
  carrying out realistic experiments with secure component loading and
  update.  } In order to achieve this goal, we will focus on the
techniques of meta-pro\-gram\-ming, program specialization and static
analysis that together support the use of \clp{} tools to analyze
\Jvml programs.  Interpretative approaches which rely on CLP tools
have been applied to analyze rather restricted versions of high-level
imperative languages~\cite{Peralta-Gallagher-Saglam-SAS98-short} and
also assembly code for PIC~\cite{HG04}, an 8-bit microprocessor.  However,
to the best of our knowledge, this is the first time the
interpretative approach has been successfully applied to a general
purpose, realistic, imperative programming language.

\subsubsection{Overview.} Fig.~\ref{fig:acc-jvm} presents a general overview of our approach. We depict an
element within a straight box to denote its use as a program and a
rounded box for data.  The whole verification process is split in
three main parts:
\begin{enumerate} 
\item \emph{Meta-programming.} We use \clp{} as a language for
  representing and manipulating \jvml{} programs. We have implemented
  an automatic translator, called \classreader, which given a set of \dotclass{}
  files $\{${\tt Class~1},$\ldots$, {\tt Class~n}$\}$ returns $P$, an
  \clp{} representation of them in \Jvmlb (a representative subset of
  \jvml{} presented in
  Sect.~\ref{sec:parser}).  Furthermore, we  also describe in Sect.~\ref{sec:spec-dynam-semant}  an interpreter in \clp{}, called  $\interpreterp$, which captures the JVM semantics. The
  interpreter has been extended in order to compute \emph{execution
    traces}, which will be very useful for reasoning about certain
  properties.
\item \emph{Partial evaluation.} The development of partial evaluation
  techniques \cite{pevalbook93} has allowed the so-called
  ``interpretative approach'' to compilation which consists in specializing an interpreter w.r.t.\ a fixed object code. 
  We have used an existing \pe{} for \clp{} in order to specialize the
  $\interpreterp$ w.r.t.\ $P$.\short{(i.e., the \clp{} representation of $\{${\tt
    Class~1},$\ldots$, {\tt Class~n}$\}$ described in 1)}  As a
  result, we obtain $\specialization$, an \clp{} residual program
  which can be seen as a decompiled and translated version of $\program$
  into \clp\ (see Sect.~\ref{sec:spec}).\short{, as we will see in Sect.~\ref{sec:spec}.}
\item \emph{Verification of Java bytecode}.  The final goal is that
  the \jvml{} program can be verified by analyzing the residual
  program $\specialization$ obtained in Step 2) above by using
  state-of-the-art {\sc analyzer}s developed for \clp{}, as we will
  see in Sect.~\ref{sec:verif}.  
 \short{This allows us to reuse advanced
  analysis engines for \clp{} without the need of defining new domains
  and analyzers for the low-level language.}
\short{, in which moreover
    it is not immediate to specify and infer global properties.}
\end{enumerate}
The resulting scheme has been implemented and incorporated in the
\ciaopp~preprocessor. Our preliminary experiments show that it is
possible to infer global properties of the computation of the residual
\clp\ programs.  We believe our proposed approach is very promising in
order to bring the analysis power of declarative languages to
low-level, imperative code such as Java bytecode.

\secbeg
\section{The Class Reader ({\sc jvml} to {\sc jvml}$_{r}$ in \clp{})}
\label{sec:parser}

\secend

\short{This section and the next one describe 
the \emph{meta-programming} phase in
Fig.~\ref{fig:acc-jvm}.
In particular, this section presents the 
\classreader.}\short{and Sect.~\ref{sec:spec-dynam-semant} presents
the \interpreterp.}

As notation, we use $\programs$ to denote \clp{} programs and
$\classes$ to denote \dotclass{} files (i.e., \Jvml classes).  The input
of our verification process is a set of 
\dotclass{} files, 
denoted as $\class_1 \ldots \class_n \in \classes$,
as specified by the Java Virtual Machine Specification
\cite{ptu:java-jvm}.
Then, the
\classreader\ 
takes $\class_1 \ldots \class_n$ and returns an \clp{} file which
contains all the information 
in $\class_1 \ldots \class_n$ 
represented in our \Jvmlb
language. \Jvmlb is a representative subset of the \Jvml language
which is able to handle: classes, interfaces, arrays, objects,
constructors, 
exceptions, method call to class and instance methods, etc. For
simplicity, some other features such as packages, concurrency and types
as float, double, long and string are
left out of the chosen subset.  
For conciseness, we use\short{the prefix \Jvmlb on $\programs$, written as}
\jvmlp{}\short{,} to make it explicit that an \clp{} program contains a \Jvmlb
representation.%
\short{We believe \Jvmlb has enough expressive
power for capturing the main features regarding analysis and
verification for our purposes.  Anyway, we expect that we could
integrate the remaining features of the \Jvml when needed without much
effort. 
We decided to make use of a subset of the \Jvml instead of using the
\Jvml itself due to two reasons. On one hand, we needed a very well
formalized language, and the original \Jvml specification was an
informal English description that was incomplete and incorrect in some
respects. On the other, it was interesting for us to restrict this
complex low-level language such as Java bytecode to a small
sub-domain in order to perform the first tests, 
and then, to be able to increment it as needed. 
}%
%
%
\short{\begin{definition}[{\sc class\_reader}]
  We define the function \classreader $:{\it\classes^{+}} \to$
    \jvmlp\ which takes a set of \dotclass{} files $\class_1
  \ldots \class_n \in \classes$ and returns an \clp{} program $\program
  \in$ \jvmlp\ which is the \emph{\clp{} representation} of
  $\class_1 \ldots \class_n$.
 \end{definition}
 The \classreader{}, implemented in
 \ciao~\cite{ciao-reference-manual-1.13-short-short}, reads the \dotclass{}
 files byte by byte and organizes and interprets them as it is
 specified in the \emph{\texttt{class} file format} specification (see
 \cite{ptu:java-jvm}).}
The differences between \Jvml and \Jvmlb are essentially the
following:
\begin{enumerate}
\item{\em Bytecode factorization.} 
  Some instructions in \jvml{} have a similar behavior and have been
  factorized in \jvmlb{} in order to have fewer
  instructions\footnote{This allows covering over 200\short{bytecode}
    instructions of \jvml{} in 54 instructions in \jvmlb{}.}.
\short{,  without affecting expressiveness.}
  This 
  makes the \jvmlb{} code easier to read (as well as the traces which
  will be discussed in Sect.~\ref{sec:spec-dynam-semant}) and the
  \interpreterp{} easier to program and maintain.

\item{\em References resolution.} The original \Jvml instructions
  contain indexes onto the \emph{constant-pool} table
  \cite{ptu:java-jvm}, a structure present in the \dotclass{} file
  which stores different kinds of data (constants, field and method
  names, descriptors, class names, etc.) and which is used in order to
  make bytecode programs as compact as possible.
  The \classreader\ removes all references to the constant-pool table
  in the bytecode instructions by replacing them with the complete
  information\short{. This can be seen as an \emph{unfolding} step whose
  purpose is} to facilitate the task of the tools which need to handle
  the bytecode later.
%
\short{\footnote{Note that
    the \pe\ can automatically perform this unfolding step. But we
    prefer to have a translator with reference resolution which can be
    used independently of our current approach (e.g., by a Java
    bytecode analyzer written in \ciao\ directly).}}
\short{ Thus, we no longer need the constant-pool table and all the required
  data are included within the \jvmlb{} class representation.
}

\short{ \item \mycomment{in \jvmlb, we do removes some information
     (to take the first difference, the magic number is not anymore in
     the \jvmlb version)}
}
\end{enumerate} 
The \ciao\ file generated by the \classreader\ contains\short{, on one hand,}
the bytecode instructions for all methods in $\class_1 \ldots
\class_n$, represented as a set of facts; and also,\short{  the other hand,} a single
fact obtained by putting together all the other information available
in the \dotclass{} files (class name, methods and fields signatures,
etc.).





\begin{figure}[t]
  \centering
  \scriptsize
  \begin{tabular}[c]{l@{}l}
    \begin{minipage}{3ex}
      \begin{alltt}
1
2
3
4
5
6
7
8
9
10
11
12
13
14
15
16
17
18
19
20
21
22
23
24
25
26
27
28
29
30
31
32
33
34
35
36
37
38
\end{alltt}
\end{minipage}
&
\begin{minipage}{0.965\linewidth}
  \begin{alltt}
class(
  className(packageName(''),shortClassName('Rational')),final(false),public(true),
  abstract(false),className(packageName('java/lang/'),shortClassName('Object')),[],
  [field(
     fieldSignature(
       fieldName(
         className(packageName(''),shortClassName('Rational')),shortFieldName(num)),
       primitiveType(int)),
     final(false),static(false),public,initialValue(undef)),
   field(
     fieldSignature(
       fieldName(
         className(packageName(''),shortClassName('Rational')),shortFieldName(den)),
         primitiveType(int)),
     final(false),static(false),public,initialValue(undef))],
  [method(
     methodSignature(
       methodName(
         className(packageName(''),shortClassName('Rational')),shortMethodName('<init>')),
       [primitiveType(int),primitiveType(int)],none),
     bytecodeMethod(3,2,0,methodId('Rational_class',1),[]),
     final(false),static(false),public),
   method(
     methodSignature(
       methodName(
         className(packageName(''),shortClassName('Rational')),shortMethodName(exp)),
       [primitiveType(int)],
       refType(classType(className(packageName(''),shortClassName('Rational'))))),
      bytecodeMethod(4,4,0,methodId('Rational_class',2),[]),
      final(false),static(false),public),
   method(
     methodSignature(
       methodName(
         className(packageName(''),shortClassName('Rational')),shortMethodName(expMain)),
       [primitiveType(int),primitiveType(int),primitiveType(int)],
       refType(classType(className(packageName(''),shortClassName('Rational'))))),
     bytecodeMethod(3,4,0,methodId('Rational_class',3),[]),
     final(false),static(true),public)]).
\end{alltt}
    \end{minipage}
  \end{tabular}
  \caption{Extract of the Program Fact Describing the Rational Class of  Running Example}
  \label{fig:exp-class}
\end{figure}

\begin{figure}[t]\scriptsize\centering%
  \begin{tabular}[c]{c}\begin{minipage}[c]{1.0\linewidth}
      \begin{alltt}
bytecode(0,2,'Rational',const(primitiveType(int),1),1).
bytecode(1,2,'Rational',istore(2),1).
bytecode(2,2,'Rational',const(primitiveType(int),1),1).
bytecode(3,2,'Rational',istore(3),1).
bytecode(4,2,'Rational',iload(1),1).
bytecode(5,2,'Rational',if0(leInt,23),3).
bytecode(8,2,'Rational',iload(2),1).
bytecode(9,2,'Rational',aload(0),1).
bytecode(10,2,'Rational',getfield(fieldSignature(
         fieldName(className(packageName(''),shortClassName('Rational')),shortFieldName(num)),
         primitiveType(int))),3).
bytecode(13,2,'Rational',ibinop(mulInt),1).
bytecode(14,2,'Rational',istore(2),1).
bytecode(15,2,'Rational',iload(3),1).
bytecode(16,2,'Rational',aload(0),1).
bytecode(17,2,'Rational',getfield(fieldSignature(
         fieldName(className(packageName(''),shortClassName('Rational')),shortFieldName(den)),
         primitiveType(int))),3).
bytecode(20,2,'Rational',ibinop(mulInt),1).
bytecode(21,2,'Rational',istore(3),1).
bytecode(22,2,'Rational',iinc(1,-1),3).
bytecode(25,2,'Rational',goto(-21),3).
bytecode(28,2,'Rational',new(className(packageName(''),shortClassName('Rational'))),3).
bytecode(31,2,'Rational',dup,1).
bytecode(32,2,'Rational',iload(2),1).
bytecode(33,2,'Rational',iload(3),1).
bytecode(34,2,'Rational',invokespecial(methodSignature(
         methodName(
            className(packageName(''),shortClassName('Rational')),
            shortMethodName('<init>')),
         [primitiveType(int),primitiveType(int)],none)),3).
bytecode(37,2,'Rational',areturn,1).

bytecode(0,3,'Rational',new(className(packageName(''),shortClassName('Rational'))),3).
bytecode(3,3,'Rational',dup,1).
bytecode(4,3,'Rational',iload(0),1).
bytecode(5,3,'Rational',iload(1),1).
bytecode(6,3,'Rational',invokespecial(methodSignature(
         methodName(
            className(packageName(''),shortClassName('Rational')),
            shortMethodName('<init>')),
         [primitiveType(int),primitiveType(int)],none)),3).
bytecode(9,3,'Rational',iload(2),1).
bytecode(10,3,'Rational',invokevirtual(methodSignature(
         methodName(
            className(packageName(''),shortClassName('Rational')),
            shortMethodName(exp)),
         [primitiveType(int)],
         refType(classType(className(packageName(''),shortClassName('Rational')))))),3).
bytecode(13,3,'Rational',areturn,1).
\end{alltt}\end{minipage}\end{tabular}%
\caption{Extract of the Bytecode facts of our Running Example}\label{fig:exp-bc}
\end{figure}

\begin{example}[running example]\label{ex:exp-java}
  Our running example considers a main Java class named
  \texttt{Rational} which represents rational numbers using two
  attributes: \texttt{num} and \texttt{den}. The class has a
  constructor, an instance method \texttt{exp} for computing the exponential of
  rational numbers w.r.t.\ a given exponent (the result is returned
  on a new rational object), and a static method \texttt{expMain}
  which given three integers, creates a new rational object using the
  first two ones as numerator and denominator, respectively, and
  invokes its \texttt{exp} method using the third argument as
  parameter. Finally, it returns the corresponding rational object. 
  This example features arithmetic operations, object creation, field
  access, and invocation of both class and instance methods. It also
  shows that our approach is not restricted to intra-procedural
  analysis. %

   In Fig.~\ref{fig:exp-class}, we show the extract of the program
  fact corresponding to class \texttt{Ra\-tion\-al}. Line
  numbers are provided for convenience but they are not part of the
  code.  The description of the field \texttt{num} appears in Lines 4-9,
 \texttt{den} in~L.10-15 and the 
  methods in~L.16-38.
  For conciseness,\short{the default constructor has been removed
    and} only methods actually used are shown.  The first
  method~(L.16-22) is a constructor that takes two integers~(L.20) as
  arguments. The second method~(L.23-30) is named \texttt{exp}~(L.26),
  it is an instance method (cf. \texttt{static(false)}~L.30)) and
  takes an integer~(L.27) as a parameter and returns an instance of
  \texttt{Rational}~(L.28). Finally, the last method~(L.31-38),
  \texttt{expMain}, is a class method (cf.
  \texttt{static(true)}~L.38), that takes as parameters three
  integers~(L.35) and returns an instance of \texttt{Rational}~(L.36).
  
  Fig.~\ref{fig:exp-bc} presents the bytecode facts corresponding to
  the methods \texttt{exp} and \texttt{expMain}.  Each fact is of the
  form \texttt{bytecode(PC,MethodID,Class,Inst,Size)}, where
  \texttt{Class} and \texttt{MethodID}, respectively, identify the
  class and the method to which the instruction \texttt{Inst} belongs.
  \texttt{PC} corresponds to the program counter and \texttt{Size} to
  the number of bytes of the instruction in order to be able to
  compute the next value of the program counter.
  The class method number 3 (i.e., \texttt{expMain}) creates first an
  instance of \texttt{Rational} (Instructions 0-6) and then invokes
  the instance method \texttt{exp} (I.9-10). The bytecode of the
  method number 2 (i.e., \texttt{exp}),
%
%
  can be divided in 3 parts. First, the initialization (I.0-3) 
  of two local variables, say $x_2$ and $x_3$,
  to 1. 
  Then, the loop body (I.4-25) first compares the exponent to 0 and, if
  it is less or equal to 0, exits the loop by jumping 23 bytes ahead
  (I.4-5).  Then, the current value of $x_2$ (\texttt{iload}) and the
  denominator (\texttt{aload} and \texttt{getfield}) are retrieved
  (I.8-10), multiplied and stored in $x_2$ (I.13-14). The same is done
  for $x_3$ with the numerator in I.15-21. Finally, the value of the
  exponent is decreased by one (I.22) and PC is decreased by 21
  (I.25) i.e., we jump back to the beginning of the loop.
  After the loop,\short{the last part of} the method creates an instance of
  \texttt{Rational}, stores the result (I.28-34), and returns this
  object (I.37).


\end{example}


\newcommand{\sn}{L}
\newcommand{\sns}{Step\_Na\-me}
\newcommand{\execute}{\textsc{execute}}
\newcommand{\traces}{Trace}
\newcommand{\stepr}{\xrightarrow{\sn}_\program}
\newcommand{\steps}{\xrightarrow{T}^*_\program}

\secbeg
\section{Specification of the Dynamic Semantics}
\secend
\label{sec:spec-dynam-semant}

(C)LP programs have been used traditionally for expressing the
semantics of both high- and low-level
languages~\cite{Peralta-Gallagher-Saglam-SAS98-short,DBLP:conf/meta/Ross88-short}.
In our approach, we express the \jvml{} semantics in \ciao.  The formal
\jvml\ specification chosen for our work is \bicolano~\cite{bicolano},
which is\short{a superset\footnote{It also includes the \texttt{tableswitch}
  and \texttt{lookupswitch} instructions.} of \jvmlb{}.  \bicolano\ is}
written with the Coq Proof Assistant~\cite{coq-short}. This allows checking
that the specification is consistent and also proving properties on
the behavior of some programs.

In the specification, a state is modeled by a 3-tuple\footnote{Both in
  Bicolano and in our implementation there is another kind of state
  for exceptions, but we have omitted it from this
  formalization for the sake of simplicity.} $\langle~Heap, Frame,
Stack\-Frame~\rangle$ which represents the machine's state where
$Heap$ represents the contents of the heap, $Frame$ represents the execution state of the current
  \emph{Method} and, $StackFrame$ is a list of frames corresponding to the call
  stack.
\short{:
\begin{itemize}
\item $Heap$ represents the contents of the heap,
\item $Frame$ represents the execution state of the current
  \emph{Method},
\item $StackFrame$ is a list of frames corresponding to the call
  stack.
\end{itemize}
}
Each frame is of the form $\langle~Method, PC, OperandStack,
LocalVar~\rangle$ and contains the stack of operands $OperandStack$
and the values of the local variables $LocalVar$ at the program point
$PC$ of the method $Method$.
\short{
As seen in section~\ref{sec:jvmlb}, \jvmlb{} is close to the language
of the official Java Virtual Machine Specification, its dynamic
semantics is therefore close to the official one. However, this
official specification is not formal and includes several problems.
Several works have been done to solve those problems.}
The definition of the dynamic semantics is based on the notion of
\emph{step}.

\begin{definition}[$step \stepr$]\label{def:steps}
  The dynamic semantics of each instruction is specified as a partial
  function $step: \textnormal{\jvmlp} \times \states \to \states
  \times \sns$ that, given a program $\program \in $ \jvmlp{} and a
  state $S \in \states$, computes the next state $S' \in \states$ and
  returns the name of the step $\sn \in \sns$.  For convenience, we
  write $S \stepr S'$ to denote $step(\program,S)=(S',\sn)$.
\end{definition}

\noindent In order to formally define our interpreter, we need to
define the following function which iterates over the steps of the
program until obtaining a final state.


\begin{definition}[$\steps$]
  Let $\steps$ be a relation on $\states$ with $S \steps S'$ iff:
  \begin{itemize}
  \item there exists a sequence of steps $\sn_1$ to $\sn_n$
    such that $S \xrightarrow{\sn_1}_\program \ldots
    \xrightarrow{\sn_n}_\program S'$,
  \item there is no state $S'' \in \states$ such that  $S' \stepr S''$, and
  \item $T \in Traces$ such that $T=[\sn_1,\ldots,\sn_n]$ is the list
    of the names of the steps.
  \end{itemize}
\end{definition}
We can now define\short{two different interpreters. One that takes
  as its only parameters a program and its arguments
  list 
  (a list of strings), and starts the execution of the \texttt{public
    static void main(java.lang.String[])} method of the first class of
  the program.  This has been implemented, but we have also defined a
  more } a general interpreter which takes as parameters a program and a
\emph{method invocation specification} (\MIS\ in the following) that
indicates: 1) the method the execution should start from, 2) the
corresponding effective parameters of the method which will often
contain logical variables or partially instantiated terms (and should
be interpreted as the set of all their instances) and 3) an initial
heap.  The interpreter relies on an \execute\ function that takes as
parameters a program $\program \in \textnormal{\jvmlp}$ and a state $S
\in \states$ and returns $(S',T)$ where $S \steps S'$.

\short{%
\begin{definition}[\execute]
  Let $\program \in \textnormal{\jvmlp}$ be a program to be executed
  and $S \in \states$ be a state. We define the execution of this
  program as
  $\execute(P,S) = (S',T)$ with $S \steps S'$.
\end{definition}
}

The following definition of \interpreterp{} computes, in addition to
the return value of the method called, also the trace which captures
the computation history.  Traces represent the semantic steps used and
therefore do not only represent instructions, as the context has also
some importance.  They allow us to distinguish, for example, for a
same instruction, the step that throws an exception from the normal
behavior. E.g.,  \texttt{invokevirtual\_step\_ok} and
\texttt{invokevirtual\_step\_\-Null\-PointerException} represent,
respectively, a normal method call and a method call on a null
reference that throws an exception. 

\begin{definition}[\interpreterp]
\label{def:interpreter}
Let $M$ be a \MIS\ that contains a method signature, the parameters
for the method and a heap, written as $M \in \MIS$.  We define a
general interpreter $\interpreterp(P,M)=(R,T)$ with
  \begin{itemize}
  \item $S=\mathit{initialState}(P,M)$, where function
    $\mathit{initialState}$ builds, from the program $P$ and the \MIS\
    $M$, a state $S \in \states$,
  \item $\execute(P,S) = (S',T)$ and
   \short{%
    \item $\mathit{finalState}(S')$ which checks that $S'$ is a valid final
      state, i.e.,   the program counter points to a
      \texttt{return} instruction and the call stack is empty, 
    }
  \item $R=\mathit{result\_of}(S')$ is the result of the execution of the
    method specified by $M$ (the value on top of the stack of the
    current frame of $S'$).
  \end{itemize}
\end{definition}
\short{%
  If the state computed by \execute\ is not a final state, then
  \interpreterp\ fails. When we can prove non failure, it means that
  the initial state built from the provided \MIS\ is guaranteed to be
  consistent.  }
%
%
This definition of \interpreterp{} returns the trace and the result of
the method but it is straightforward to modify the definitions of
\interpreterp{} and \execute{}\short{(and the corresponding code)} to return
less information or to add more. This gives more flexibility to our
interpretative approach when compared to direct compilation: for
example, if needed, we can return in an additional argument a list
containing the information about each state which we would like to
\emph{observe} in order to prove properties which may require a deeper
inspection of execution states.




\secbeg
\section{Automatic Generation of Residual Programs}\label{sec:spec}
\secend

Partial evaluation (PE) \cite{pevalbook93} is a semantics-based
program optimization technique which has been deeply investigated
within different programming paradigms.
\short{ and applied to a wide variety
of languages.}
The main purpose of PE is to specialize
a given program w.r.t.\ the \emph{static data}, i.e., the part of its
input data which is known---hence it is also known as \emph{program
  specialization}.  The partially evaluated (or residual) program will
be (hopefully) executed more efficiently since those computations that
depend only on the static data are performed once and for all at PE
time. We use the partial evaluator for \clp{}
programs of~\cite{ai-with-specs-sas06-short}\short{written in \ciao\
and} which is part of \ciaopp. Here, we represent it as a function
\pe$:\programs \times \data \to \programs$ which, for a given program
$\program \in \programs$ and static data $S \in \data$, returns a
residual program $\program_{S} \in \programs$ which is a
\emph{specialization} \cite{pevalbook93} of $\program$ w.r.t.\
$S$.


The development of PE, program specialization and
related techniques
\cite{Futamura:71:54,pevalbook93,gallagher86-short}
has led to 
an alternative 
approach to compilation (known as the first Futamura projection) based
on specializing an interpreter with respect to a fixed object program.
The success of the application of the technique involves eliminating
the overhead of parsing the program, fetching instructions, etc., and
leading to a residual program whose operations mimic those of the
object program. This can also be seen as a translation of the object
program into another programming language, in our case \ciao.  The
\emph{residual} program is ready now to be, for instance, efficiently
executed in such language or, as in our case, accurately analyzed by
tools for the language in which it has been translated.
\short{German:In the \clp{} context, this interpretative approach has
  been applied to analyze high-level imperative languages
  \cite{Peralta-Gallagher-Saglam-SAS98-short} and also the PIC
  processor \cite{HG04} by relying on CLP tools. We are not aware of
  any attempt to use the interpretative approach to the automatic
  verification of Java bytecode.  However, this is the first time the
  approach has been applied in the context of Java bytecode. The new
  challenges it introduces include the existence of a stack...
  complete!}
The application of this interpretative approach to compilation 
within our framework consists in partially evaluating
the $\interpreter$ w.r.t.\short{a program 
} $\program=\mbox{\classreader}(\class_1,\ldots,\class_n)$ and a \MIS.
\short{ (see Def.~\ref{def:interpreter} above). This results
in a residual \clp{} program, $\specialization$.  One can think
  of $\specialization$ as an \clp{} translation of
  $\class_1,\ldots,\class_n$, emulating the part of its semantics
  captured by the $\interpreter$.  }

\begin{definition}[\clp{} residual program]
  Let $\interpreter \in \programs$ be a \jvmlb{} interpreter, $M \in
  \MIS$ and $\class_1,\ldots,\class_n \in \classes$ be a set of
  classes. The \emph{\clp{} residual program}, $\specialization$, for
  $\interpreter$ w.r.t.  $\class_1,\ldots,\class_n$ and $M$ is defined
  as
  $\specialization$=\pe($\interpreter,(\textnormal{\classreader}(\class_1,\ldots,$
  $\class_n),M))$.
\end{definition}
Note that, instead of using the interpretative approach, we could have
implemented a compiler from Java bytecode to \clp.  However, we
believe that the interpretative approach has at least the following
advantages: 1) more flexible, in the sense that it is easy to modify
the interpreter in order to observe new properties of interest, see
Sect.~\ref{sec:spec-dynam-semant}, 2) easier to trust, in the sense
that it is rather difficult to prove (or trust) that the compiler
preserves the program semantics and, it is also complicated to
explicitly specify what the semantics used is, 3) easier to maintain,
new changes in the JVM semantics can be easily reflected in the
interpreter by modifying (or adding) a proper ``step'' definition, and
4) easier to implement, provided a powerful partial evaluator for
\clp{}\short{programs} is available.


\begin{example}[residual programs]\label{ex:pe_notrace}
  We now want to partially evaluate our implementation of the
  interpreter which does not output the trace (see
  Sect.~\ref{sec:spec-dynam-semant}) w.r.t.\ the  bytecode method \texttt{expMain}
in Ex.~\ref{ex:exp-java}, an empty heap and three free variables as parameters.
The size of the program to be partially evaluated (i.e., interpreter)
is 86,326 bytes (2,240 lines) while the size of the data (i.e.,
bytecode representation) is 16,677 bytes (101 lines) of \jvmlb{}.  The
  partial evaluator has different options for tuning the level of
  specialization.  For this example, we have used local and global
  control strategies based on \emph{homeomorphic embedding}
  (see~\cite{Leuschel:SAS98-short}).\short{for details).}
\short{In particular, the so-called local control decides
  when to stop derivations and the global control when to generalize a
  new term resulting from a previous unfolding. For this example, we
  have used the local control strategy based on \emph{homeomorphic
    embedding} which is described in~\cite{lopstr04-unfolding-short}.
  For the global control, we have also used homeomorphic embedding in
  order to flag when generalization is required.
}

We show in Fig.~\ref{fig:res-exp-nt} the residual program resulting of
such automatic PE.  The parameters \texttt{A},
\texttt{B} and \texttt{C} of \texttt{expMain/5} represent the
numerator, denominator and exponent, respectively. The fourth and fifth
parameters represent, respectively, the top of the stack and the heap where the method result (i.e., an object
of type \texttt{Rational} in the bytecode) will be returned. In
particular, the result corresponds to the second element,
\texttt{ref(loc(2))}, in the heap. Note that this object is
represented in our LP program as a list of two atoms, the first one
corresponds to attribute \texttt{num} and the second one to
\texttt{den}. The first two rules for \texttt{expMain/5} are the base
cases for exponents $\tt C=0$ and $\tt C=1$, respectively. The third
rule, for $\tt C>1$, uses an auxiliary recursive predicate
\texttt{execute/6} which computes $\tt A^{C+1}$ and $\tt B^{C+1}$ and
returns the result in the second element of the heap. It should be
noted that our PE tool has done a very good job by transforming a
rather large interpreter into a small residual program (where all the
interpretation overhead has been removed).  The most relevant point to
notice about the residual program is that we have converted low level
jumps into a recursive behavior and achieved a very satisfactory
translation from the Java bytecode method \texttt{expMain}. Indeed,
it is not very different from the \ciao\ version one could have
written by hand, provided that we need to store the result in the
fifth argument of predicate \texttt{expMain/5} as an object in the
heap, using the corresponding syntax.



\begin{figure}[t]
\centering
\small
\begin{minipage}{11cm}
\begin{alltt}
expMain(A,B,C,ref(loc(2)),heap([[num(int(A)),num(int(B))],
            [num(int(1)),num(int(1))]])) :- C=<0 .
expMain(A,B,C,ref(loc(2)),heap([[num(int(A)),num(int(B))],
            [num(int(A)),num(int(B))]])) :- C>0, F is C-1, F=<0 .
expMain(A,B,C,D,E) :- C>0, H is C-1, H>0, I is A*A, 
            J is B*B, K is H-1, execute(A,B,K,I,J,E,D) .

execute(A,B,C,D,E,heap([[num(int(A)),num(int(B))],
            [num(int(D)),num(int(E))]]),ref(loc(2))) :- C=<0 .
execute(A,B,C,D,E,G,L) :- C>0, N is D*A, O is E*B, P is C-1,
            execute(A,B,P,N,O,G,L) .
\end{alltt}
\end{minipage}
\caption{Residual Exponential Program without Trace}\label{fig:res-exp-nt}
\vspace*{-0.5cm}
\end{figure}

%

While the above \clp\ program\short{which computes the same result as
  a bytecode method} can be of a lot of interest when reasoning about
functional properties of the code, it is also of great importance to
augment the interpreter with an additional argument which computes a
trace (see Def.~\ref{def:interpreter}) in order to capture the
computation history.  The residual program which computes execution
traces is \texttt{expMain/4}, which on success contains in the fourth
argument the execution trace at the level of Java bytecode (rather
than the top of the stack and the heap). Below, we show the recursive
rule of predicate \texttt{execute/8} whose last argument represents
the trace (and corresponds to the second rule of \texttt{execute/7}
without trace in Fig.~\ref{fig:res-exp-nt}): {\small\allttbeg
\begin{alltt}
execute(B,C,D,E,F,G,I,[goto_step_ok,iload_step,if0_step_continue,
         iload_step,aload_step_ok,getfield_step_ok,ibinop_step_ok,
         istore_step_ok,iload_step,aload_step_ok,getfield_step_ok,
         ibinop_step_ok,istore_step_ok,iinc_step|H]) :-
     D>0, I is E*B, J is F*C, K is D-1, execute(B,C,K,I,J,G,I,H) .
\end{alltt}\allttend}
\noindent As we will see in the next section, this trace will allow observing a good
number of interesting properties about the program.

\end{example}


\secbeg
\section{Verification of Java Bytecode Using \clp{} Analysis Tools}\label{sec:verif}
\secend

Having obtained an \clp{} representation of a Java bytecode program,
the next task is to use existing analysis tools for \clp{} in order to
infer and verify properties about the original bytecode program.  
\short{The
analysis tools we use are based on \emph{abstract
interpretation} \cite{Cousot77-short} and are part of the \ciaopp\
system~\cite{ciaopp-sas03-journal-scp-short-short}.}
We now recall some basic notions on \emph{abstract
interpretation}~\cite{Cousot77-short}. \emph{Abstract
interpretation} provides a general
formal framework for computing safe approximations
of program beha\-vi\-our.
%
In this framework, 
programs are interpreted using \emph{abstract values} instead of
\emph{concrete values}.
%
%
An abstract
value is a finite representation of a, possibly infinite, set of
concrete values in the concrete domain $D$.  
The set of all possible abstract values constitutes the \emph{abstract
  domain}, denoted $D_\alpha$, which is usually a complete lattice or
cpo which is ascending chain finite.  Abstract values and sets of concrete
values are related by an {\em abstraction} function $\alpha:
2^{D}\rightarrow \dom$, and a {\em concretization} function $\gamma:
\dom\rightarrow 2^{D}$.  
  The concrete and abstract domains must be related in such a way that
  the following condition holds~\cite{Cousot77-short}: $\forall x\in 2^D:~
  \gamma(\alpha(x)) \supseteq x \mbox{~~~and~~~} \forall y\in
  \dom:~ \alpha(\gamma(y)) = y$.
\noindent
In general, the comparison in $\dom$, written $\sqsubseteq$, is
induced by $\subseteq$ and $\alpha$.

 We rely on a generic analysis algorithm (in the style of
\cite{ciaopp-sas03-journal-scp-short-short}) defined as a function {\sc
analyzer}$:\programs \times \aatom \times \adom \to \acert$
 which takes a program $\program
\in \programs$, an abstract domain $\dom \in \adom$ and a set of 
abstract atoms $\atom \in \aatom$ which are descriptions of the
entries (or calling modes) into the program and returns $\cert \in
\acert$.
Correctness of analysis ensures that $\cert$ safely approximates the
semantics of $\program$. We denote that $\atom$ and $\cert$ are
abstract semantic values in $\dom$ by using the same subscript
$\alpha$. \short{($\sqbrack{\program})$, i.e., $\sqbrack{P}\in
\gamma(\sqbrack{P}_\alpha)$.}

In order to verify the program, the user has to provide the intended
semantics $\inten$ (or\short{program} specification) as a semantic value in $
\dom$ in terms of \emph{assertions} (these are linguistic
constructions which allow expressing properties of programs)
\cite{assert-lang-disciplbook-short-short}.  This intended semantics
embodies the requirements as an expression of the user's expectations.
The \emph{verifier} has to compare the (actual) inferred semantics
$\cert$ w.r.t.\ $\inten$.\short{\footnote{Comparison between actual and
intended semantics of the program is most easily done in the same
domain, since then the operators on the abstract lattice, that are
typically already defined in the analyzer, can be used to perform this
comparison.}}  We use the \emph{abstract
interpretation}-based verifier
integrated in \ciaopp. It is dealt here as a function
\verifier$:\programs \times \aatom \times \adom \times \aint \to
boolean$ which for a given program $\program \in \programs$, a set of
abstract atoms $\atom \in \aatom$, an abstract domain $\dom \in \adom$
and an intended semantics $\inten$ in $\dom$ succeeds if the
approximation computed by {\sc
analyzer}($\program,\atom,\dom$)=$\cert$ entails that $\program$
satisfies $\inten$, i.e., $\cert\sqsubseteq \inten$.

\short{In our setting the program to be analyzed and verified is a
  \clp{} residual program that emulates and represents the Java
  bytecode.  Clearly from the point of view of the verifier, the fact
  that it is a residual program is not relevant.}

\begin{definition}[verified bytecode]\label{def:verif}
  Let $\specialization \in \programs$ be an \clp{} residual program
  for $\interpreterp$ w.r.t.  $\class_1,\ldots,\class_n \in \classes$
  and  $M \in \MIS$ (see
  Def.~\ref{def:interpreter}).  Let $\dom \in \adom$ be an abstract
  domain, $\atom \in \aatom$ be a set of abstract atoms and $\inten
  \in \dom$ be the abstract intended semantics of
  $\specialization$. We say that $(\class_1,\ldots,\class_n,M)$ is
  verified w.r.t.\ $\inten$ in $\adom$ if
  $\mbox{\verifier}(\specialization$, $\atom, \dom, \inten)$ succeeds.
\end{definition}
In principle, any of the considerable number of abstract domains
developed for \emph{abstract
interpretation} of logic programs can be applied
to residual programs, as well as to any other
program. 
\short{Also, the implemented systems such as \ciaopp.}
In addition, arguably,
analysis of logic programs is inherently simpler than that of Java
bytecode since the bytecode programs decompiled into logic programs no
longer contain an operand stack for arithmetic and execution flow is
transformed from jumps (since loops in the Java program are compiled
into conditional and unconditional jumps) into recursion.
\short{In the next sections, we briefly illustrate the kind of
properties we can verify about them with the \ciaopp\ system.}

\subsecbeg
\subsection{Run-Time Error Freeness Analysis}\label{sec:runtime}
\subsecend




The use of objects in Ex.~\ref{ex:exp-java} could in
principle issue exceptions of type \texttt{Null\-Poin\-ter\-Ex\-cep\-tion}.
Clearly, the execution of the \texttt{expMain} method will not produce
any exception, as the unique object used is created within the method.
However, the JVM is unaware of this and has to perform the
corresponding run-time test.  We illustrate that by using our approach
we can statically verify that the previous code cannot issue such an
exception (nor any other kind of run-time error).

First, we proceed to specify in \ciao\ the property ``{\tt
  goodtrace}'' which encodes the fact that a bytecode program is
run-time error free in the sense that its execution does not issue
\texttt{NullPointerException} nor any other kind of run-time error
(e.g., \texttt{Array\-In\-dex\-Out\-Of\-Bounds\-Exception}, etc). As
this property is not predefined in \ciao, we declare it as a regular
type using the {\tt regtype} declarations in
\ciaopp.
Formally, we define this property as a \emph{regular
    unary logic} program, see \cite{FruewirthShapiroVardiYardeni91}.
The following regular type {\tt goodtrace} defines this notion of
safety for our example (for conciseness, we omit the bytecode
instructions which do not appear in our program):



{\scriptsize\allttbeg
\begin{alltt}
:- regtype goodtrace/1.
goodtrace(T) :- list(T,goodstep).

:- regtype goodstep/1.
goodstep(iinc_step).         goodstep(aload_step_ok).        goodstep(invokevirtual_step_ok).         
goodstep(iload_step).        goodstep(if0_step_jump).        goodstep(invokestatic_step_ok).                         
goodstep(normal_end).        goodstep(const_step_ok).        goodstep(if0_step_continue).          
goodstep(new_step_ok).       goodstep(return_step_ok).       goodstep(if_icmp_step_jump).                          
goodstep(pop_step_ok).       goodstep(astore_step_ok).       goodstep(putfield_step_ok).            
goodstep(dup_step_ok).       goodstep(istore_step_ok).       goodstep(getfield_step_ok).            
goodstep(goto_step_ok).      goodstep(ibinop_step_ok).       goodstep(if_icmp_step_continue).         
goodstep(areturn_step_ok).   goodstep(invokespecial_step_here_ok).
\end{alltt}\allttend}
\noindent Next, the version with traces of the residual program in Fig.~\ref{fig:res-exp-nt} is
extended with the following assertions: \subsecbeg
{\small\allttbeg
\begin{alltt}
:- entry expMain(Num,Den,Exp,Trace):(num(Num),num(Den),num(Exp),var(Trace)).
:- check success expMain(Num,Den,Exp,Trace) => goodtrace(Trace).
\end{alltt}\allttend}
\noindent The entry assertion describes the valid external queries to
predicate \texttt{expMain/4}\short{(i.e, it plays the role of abstract atom in
Def.~\ref{def:verif})}, where the first three parameters are 
of type \texttt{num} and the fourth one is a variable.
 We use the ``\texttt{success}'' assertion as a
way to provide a partial specification of the program. It should be
interpreted as: for all calls to \texttt{expMain(Num,Den,Exp,Trace)}, if the call
succeeds, then \texttt{Trace} must be a \texttt{goodtrace}.

%

Finally, we use \ciaopp\ to perform regular type analysis using\short{, for
example,} the \emph{eterms} domain \cite{eterms-sas02-short}. This
allows computing safe approximations of the success states of all
predicates. After this, \ciaopp~performs compile-time
checking of the
\texttt{success} assertion above, comparing it with the assertions
inferred by the analysis, and produces as output the following assertion:
{\small\allttend
\begin{alltt}
:- checked success expMain(Num,Den,Exp,Trace) => goodtrace(Trace).
\end{alltt}\allttend}
\noindent Thus, the provided assertion has been  \emph{validated}
(marked as \texttt{checked}).
\short{ When all \texttt{check}
 assertions (in this case only one) have been transformed into this
 \texttt{checked} status, the program has been \emph{verified}.
}

\subsecbeg
\subsection{Cost Analysis and Termination}\label{sec:cost}
\subsecend

As mentioned before, \emph{abstract
interpretation}-based program analysis
techniques allow inferring very rich information including also
resource-related issues.
\short{ (e.g., we can prove that the code will not
compute for more than a given amount of time, or that it will not take
up an amount of memory or other resources above a certain threshold).}
For example, \ciaopp\ can compute upper and lower bounds on the number
of execution steps required by the
computation~\cite{ciaopp-sas03-journal-scp-short-short,low-bou-sas94-short-short}.
Such bounds are expressed as functions on the sizes of the input
arguments.  Various metrics are used for the ``size'' of an input,
such as list-length, term-size, term-depth, integer-value, etc. Types,
modes, and size measures are first automatically inferred by the
analyzers and then used in the size and cost analysis.

Let us illustrate the cost analysis in \ciaopp\ on our running
example. We consider a slightly modified version of the residual
program in Fig.~\ref{fig:res-exp-nt} in which we have eliminated the accumulating
parameter due to a current limitation of the cost analysis in
\ciaopp.
\short{This limitation is currently being overcome by the new
extension of the cost analysis. Moreover, there exists an
automatic transformation to remove this parameter \cite{KGK01}.}
The cost analysis can then infer the following property of the
recursive predicate \texttt{execute/5} (and a similar one of \texttt{expMain/4}) 
using the same entry assertion as in  Sect.~\ref{sec:runtime}:
{\small\allttbeg
\begin{alltt}
:- true pred execute(A,B,C,D,E): (num(A),num(B),num(C),var(D),var(E))
        => ( num(A), num(B), num(C), num(D), num(E), 
             size_ub(A,int(A)), size_ub(B,int(B)), size_ub(C,int(C)), 
             size_ub(D,expMain(int(A),int(C)+1)+int(A)), 
             size_ub(E,expMain(int(B),int(C)+1)+int(B)) )
         + steps_ub(int(C)+1).
\end{alltt}\allttend}
\noindent which states that \texttt{execute/5} is called in this program with
the first three parameters being of type \texttt{num} (i.e., bound to
numbers) and two variables. The part of the assertion after the
\texttt{=>} symbol indicates that on success of the predicate all five
parameters are bound to numbers.  This is used by the cost analysis in order to
set the integer-value as size-metric for all five arguments.
The first three arguments are input to the procedure and thus their
size (value) is fixed. The last two arguments are output and their
size (value) is a function on the value of (some of) the first three
arguments. The upper bound computed by the analysis for $D$ (i.e., the
fourth argument) is $A^{C+1}+A$. Note that this is a correct upper
bound, though the most accurate one is indeed $A^{C+1}$.  A similar
situation occurs with the upper bound for the fifth argument ($E$).
Finally, the part of the assertion after the \texttt{+} symbol
indicates that an upper bound on the number of execution steps is
$C+1$, which corresponds to a linear algorithmic complexity. 
This is indeed the most accurate upper bound possible, since
predicate \texttt{execute/5} is called 
$C+1$ times
until $C$ becomes zero.
Note that, in this case, we do not mean the number of JVM steps in Def.~\ref{def:steps}, but the
number of computational steps. 

\short{Due to a current limitation of this analysis, the residual
program must not have accumulating parameters. Unfortunately, the
partial evaluator integrated into \ciaopp\ generates residual program
with accumulating parameters. Although the most studied problem is
usually the opposite, that is to say, adding accumulative parameters
in order to reduce computation time by allowing the compiler not to
have to allocate a new execution frame for the recursion and using the
current one, a solution to remove this parameter is proposed in
[18]. Because of a lack of time, we have not yet implemented this in
CiaoPP. Fig. 2.5 is a modified-by-hand version of the residual
programing order to get rid of the accumulating parameters and to
simplify the state to make the program easier to read.  }


\short{
\subsecbeg
\subsection{Termination Analysis}
\subsecend



Program termination is obviously a desirable property in many
contexts. Unfortunately, and as it is well known, this is an
undecidable property, and therefore we can only expect termination
analysis to compute approximate results. In spite of this, powerful
static analyzers are available which can ensure termination for an
important subset of terminating programs.
In the termination analysis area, it can be argued that the state of
the art in \clp{} is more advanced than that in imperative
programming. 
}

\ciaopp's termination analysis relies on the cost
analysis described in the previous section. In particular, it is able
to prove termination of a program provided it obtains a non-infinite
upper bound of its cost.  
Following the example of Sect.~\ref{sec:cost}, \ciaopp\ is able to turn into \texttt{checked} status
 the following assertion (and the similar one for \texttt{expMain/4}):
 ``{\tt :- check comp execute(A,B,C,D,E) + terminates}''.
which ensures that the execution of the recursive 
predicate always terminates w.r.t.\ the previous entry.

\short{{\small\allttbeg
\begin{alltt}
:- check comp execute(A,B,C,D,E) + terminates.
\end{alltt}\allttend}}

\short{There are other well-known termination analysis systems for \clp{}
like TerminWeb~\cite{CT:JLP} and cTi~\cite{Mesnard:jicslp96-short}.
Either of these systems can be used in order to prove termination of
the residual \clp{} program in the previous section.}
\short{ The
argument for proving termination of all calls satisfying the entry
declaration above is as follows.  Non-termination can only occur in
loops. If (1) we can find an argument whose size decreases in every
iteration of the loop w.r.t.  some norm which assigns values always
greater or equal than zero to any term, and (2) the program is
\emph{rigid} w.r.t. the size of the corresponding argument (all
instances of the term have the same size) and, hence, the program
terminates.  In Example~\ref{ex:pe_notrace}, the only loop we have is
for predicate \texttt{execute\_4\_1/6}.  We can conclude termination
by reasoning on the last argument. This argument can be inferred to be
bound to an integer for all computations originating from the entry
assertion above. Since in the recursive path this last argument is
decreased before making the recursive call, the program is guaranteed
to terminate.
}


\secbeg
\section{Experiments and Discussion}
\secend


We have implemented and performed a preliminary experimental
evaluation of our framework within the \ciaopp{}
preprocessor~\cite{ciaopp-sas03-journal-scp-short-short}, where we have
available a partial evaluator and a generic analysis engine with a
good number of abstract domains, including the ones illustrated in the
previous section.  Our interpretative approach has required the
implementation in \ciao\ of two new packages: the \classreader\ (1141
lines of code) which parses the \texttt{.class} files into \ciao{} and the
\interpreterp\ interpreter for the \jvmlb{} (3216 lines). 
\short{ The
\classreader{} 
consists of a set of auxiliary modules in charge
of 
the representation of the translations (e.g., the method descriptor
representation), the module which parses each individual
\texttt{.class} file, and the program loader which generates the
\clp{} program.  The code of the interpreter can be structured in four
main parts: (1) the definition of the \jvmlb{} with the access
functions to the source programs, (2) the representation of the
domains as the heap, the \jvmlb{} types or the machine states, (3) the
implementation of the dynamic semantics which uses these domains and,
(4) the interface with the partial evaluator.  All the above tools,}
These tools, together with a collection of examples, are\short{publicly} available at:
{\small\verb+http://cliplab.org/Systems/jvm-by-pe+}.

\short{The generic analyzer allows inferring very rich information
about \clp{} programs, including data structure shape (with pointer
sharing), bounds on data structure sizes, and other operational
variable instantiation properties, as well as (global)
\emph{procedure-level} properties such as determinacy, termination,
non-failure, and bounds on resource consumption (time or space cost).
This work attempts to transfer such advanced features available in
\clp{} analysis to the verification of Java bytecode.  With this aim,
we first partially evaluate a Java bytecode interpreter in \clp\
w.r.t. (an \clp\ representation of) a Java bytecode and then analyze
the residual program using such \clp\ analysis tools. Our examples
show that we are able to reason about non-trivial properties of Java
bytecode programs such as termination and run-time error freeness.  }

 Table~1  studies two crucial points for the practicality of our
proposal: the size of the residual program and the relative efficiency
of the full transformation+analysis process.  As mentioned before, the
algorithms are parametric w.r.t.\ the abstract domain. In our
experiments we use {\em eterms}, an abstract domain based on regular types, that is very
useful for reasoning about functional properties of the code,
run-time errors, etc., which are crucial aspects for the safety of
the Java bytecode.  The system is implemented in Ciao 1.13
\cite{ciao-reference-manual-1.13-short-short} with compilation to WAM bytecode.
The experiments have been performed on an Intel P4 Xeon 2\ GHz with
4\ GB of RAM, running GNU Linux FC-2, 2.6.9.

The input ``program'' to be partially evaluated is the \interpreterp\
interpreter in all the examples.  Then, the first group of columns
\textbf{Bytecode} shows information about the input ``data'' to the
partial evaluator, i.e., about the \dotclass{} files. The columns
\textbf{Class} and \textbf{Size} show the names of the classes used
for the experiments and their sizes in bytes, respectively. The second
column \textbf{Method} refers to the name of the method within each
class which is going to form the \MIS, i.e., to be the starting point
for PE and context-sensitive program analysis. We use
a set of classical algorithms as benchmarks.  The first 9 methods
belong to programs with iterations and static methods but without
object-oriented features, where \textbf{mod}, \textbf{fact},
\textbf{gcd} and \textbf{lcm}, compute respectively the modulo,
factorial, greatest-common-divisor and least-common-multiple (two
versions); the \textbf{Combinatory} class has different methods for
computing the number of selections of subsets given a set of elements
for every ordering/repetition combination. 
The next two benchmarks,
\textbf{LinearSearch} and \textbf{BinarySearch}, deal with arrays and
correspond to the classic linear and binary search algorithms. 
Finally, the last four benchmarks
correspond to programs which make extensive use of 
object-oriented features such as  instance method invocation, field
accessing and setting, object creation and initialization, etc.

The information about the ``output'' of the PE process
appears in the second group of columns, \textbf{Residual}. The columns
\textbf{Size} and \textbf{NUnfs} show the size in bytes of each
residual program and the number of unfolding steps performed by the
partial evaluator to generate it, respectively.  We can observe that
the partial evaluator has done a good job in all examples by
transforming a rather large interpreter (86,326 bytes) in relatively
small residual programs. The sizes range from 317 bytes for
\textbf{m2} (99.4\% reduction) to 4.911 for \textbf{Lcm2} (83.6 \%).
The number of required unfolding steps explains the high PE times, as
we discuss below. A relevant point to note is that, for most programs,
the size of the \clp\ translation is larger than the original
bytecode. This can be justified by the fact that the resulting program
does not only represent the bytecode program but it also makes
explicit some internal machinery of the JVM. This is the case, for
instance, of the exception handling. As there are no \ciao\ exceptions
in the residual program, the implicit exceptions in \jvml\ have been
made explicit in \clp{}. Furthermore, the Java bytecode has been
designed to be really compact, while the \clp{} version has been
designed to be easier to read by human beings and contains type
information that must be inferred on the \jvml. It should not be
difficult to reduce the size of the residual bytecode if so required
by, for example, simply using short identifiers.

The final part of the table provides the times for performing the
transformations and the analysis process.  Execution times are given
in milliseconds and measure \emph{runtime}.  They are computed as the
arithmetic mean of five runs. For each benchmark, \textbf{Trans},
\textbf{PE} and \textbf{Ana} are the times for executing the
\classreader, the partial evaluator and the analyzer,
respectively. The column \textbf{Total} accumulates all the previous
times. We can observe that most of the time is due to the partial
evaluation phase (and this time is directly related to the number of
unfolding steps performed). This is to be expected because the
specialization of a large program (i.e., the interpreter) requires to
perform many unfolding steps in all the examples (ranging from 14.117
steps for \textbf{search} in \textbf{BinarySearch} to 527 for
\textbf{m2}), plus many additional generalization steps which are not
shown in the table. The analysis time is then relatively low, as the
residual programs to be analyzed are significantly smaller than the
program to be partially evaluated. 

\begin{table}[t]
\centering
\begin{tabular}{||l|c|c||c|c||c|c|c|c||}\hline\hline
 \multicolumn{3}{||c||}{\textbf{Bytecode}} 
 & \multicolumn{2}{|c||}{\textbf{Residual}} 
 &\multicolumn{4}{|c||}{\textbf{Times (ms)}}  
 \\ \hline  
\textbf{Class} & \textbf{Size} & \textbf{Method} & \textbf{Size} & \textbf{NUnfs} &\textbf{Trans} & \textbf{PE} & \textbf{Ana} & \textbf{Total}\\ \hline \hline \hline
Mod & 314 & mod & 956  & 1645 & 18 & 1244 & 59 & 1322 
\\ \hline
Fact & 324 & fact & 1007  & 1537 & 19 & 1432 & 74 & 1525 
\\ \hline
Gcd & 265 & gcd & 940  & 1273 & 18 & 1160 & 125 & 1303 
\\ \hline
Lcm & 299 & lcm & 2260  & 4025 & 21 & 5832 & 817 & 6670 
\\ \hline
Lcm2 & 547 & lcm2 & 4911  & 3724 & 26 & 3963 & 1185 & 5174 
\\ \hline
Combinatory & 703 & varNoRep & 1314  & 1503 & 32 & 1837 & 87 & 1955 
\\ \hline
Combinatory & 703 & combNoRep & 2177  & 2491 & 34 & 3676 & 150 & 3860 
\\ \hline
Combinatory & 703 & combRep & 2151  & 3033 & 29 & 5331 & 950 & 6310 
\\ \hline
Combinatory & 703 & perm & 1022  & 1256 & 29 & 1234 & 65 & 1328 
\\ \hline
LinearSearch & 318 & search & 3114  & 8832 & 22 & 45228 & 296 & 45546 
\\ \hline
BinarySearch & 412 & search & 3670  & 14117 & 23 & 72945 & 313 & 73282 
\\ \hline
Np & 387 & m2 & 317  & 527 & 20 & 502 & 12 & 534 
\\ \hline
ExpFact & 890 & main & 2266  & 8353 & 35 & 23773 & 95 & 23903 
\\ \hline
Rational & 559 & expMain & 3131  & 6613 & 31 & 13692 & 16 & 13739 
\\ \hline
Date & 602 & forward & 11046  & 26982 & 36 & 80960 & 218 & 81213 
\\ \hline
\end{tabular}
\label{tab:table}
\caption{Sizes of residual programs and transformation and analysis times}
\vspace*{-1cm}
\end{table}

As for future work, we plan to obtain accurate bounds on resource
consumption by considering the traces that the residual program
contains and the concrete cost of each bytecode instruction. Also, we
are  in the process of studying the scalability of our approach to
the verification of larger Java bytecode programs.  
\short{The analysis tools
in \ciaopp\ are designed with support for incrementality and
modularity. We hope that these features will facilitate the
scalability of our approach.
}
We also plan to exploit the advanced features of the partial evaluator
which integrates\short{the technique of \emph} abstract
interpretation~\cite{ai-with-specs-sas06-short} in order to handle recursion.

{\small
\subsecbeg
\subsubsection*{Acknowledgments}
\subsecend This work was funded in part by the Information Society
Technologies program of the European Commission, Future and Emerging
Technologies under the IST-15905 {\em MOBIUS} project, by the Spanish
Ministry (TIN-2005-09207 {\em MERIT}), and the Madrid Regional
Government (S-0505/TIC/0407 \emph{PROMESAS}). The authors would like
to thank David Pichardie and Samir Genaim for useful discussions on
the \bicolano\ JVM specification and on termination analysis,
respectively.
}


\bibliographystyle{plain}

\comment{

}

\end{document}